\documentclass[10pt]{article}
\usepackage{graphicx}
\usepackage{lineno}
\usepackage{url}

\def\Title#1{\begin{center} {\Large #1 } \end{center}}
\def\Author#1{\begin{center}{ \sc #1} \end{center}}
\def\Address#1{\begin{center}{ \it #1} \end{center}}

\newcommand\pubblock{\rightline{\begin{tabular}{l} Proceedings of the Fifth Annual LHCP\\ \pubnumber\\
         \pubdate  \end{tabular}}}

\newenvironment{Abstract}{\begin{quotation} \begin{center} 
             \large ABSTRACT \end{center}\bigskip 
      \begin{center}\begin{large}}{\end{large}\end{center} \end{quotation}}

\newenvironment{Presented}{\begin{quotation} \begin{center} 
             PRESENTED AT\end{center}\bigskip 
      \begin{center}\begin{large}}{\end{large}\end{center} \end{quotation}}

\def\beq{\begin{equation}}
\def\eeq#1{\label{#1}\end{equation}}
\def\eeqn{\end{equation}}

\def\beqa{\begin{eqnarray}}
\def\eeqa#1{\label{#1}\end{eqnarray}}
\def\eeqan{\end{eqnarray}}

\let\bar=\overbar

\def\Dslash{\not{\hbox{\kern-4pt $D$}}}
\def\dslash{\not{\hbox{\kern-2pt $\del$}}}

\def\msb{{\bar{\ssstyle M \kern -1pt S}}}

\usepackage{color}

\newcommand\pubnumber{ ATL-PHYS-PROC-2017-106 }
\newcommand\pubdate{\today}

\def\affiliation{
On behalf of the ATLAS Collaboration, \\
Laboratoire de Physique de Clermont-Ferrand (LPC)\\
Universit\'{e} Clermont Auvergne, CNRS/IN2P3 \\
Clermont-Ferrand, France}

\begin{document}

% large size for the first page
\large
\begin{titlepage}
\pubblock

%% Change the title, name, abstract
%% Title 
\vfill
\Title{  Searches for rare and exotic Higgs decays with ATLAS  }
\vfill

%  if you need to add the support use this, fill the \support definition above. 
%   \Author{ FIRSTNAME LASTNAME \support }
\Author{ Marija Marjanovi\'c  }
\Address{\affiliation}
\vfill
\begin{Abstract}

Searches for rare and exotic Higgs decays using proton-proton collision 
data with the center-of-mass energy of 8 TeV and 13 TeV collected with the ATLAS detector are 
presented. Various final states are considered. 
No significant deviations from the Standard Model expectations are found. 
The results are interpreted in different Beyond Standard Model theories.

\end{Abstract}
\vfill

% DO NOT CHANGE 
\begin{Presented}
The Fifth Annual Conference\\
 on Large Hadron Collider Physics \\
Shanghai Jiao Tong University, Shanghai, China\\ 
May 15-20, 2017
\end{Presented}
\vfill
\end{titlepage}
\def\thefootnote{\fnsymbol{footnote}}
\setcounter{footnote}{0}
%

% normal size for the rest
\normalsize 

%% Your paper should be entered below. 

\section{Introduction}

Standard Model (SM) of particle physics predicts many possible decay modes for the Higgs boson.  
Up to now Higgs boson was observed in many of them \cite{Aad:2012tfa,Chatrchyan:2012ufa}. 
Some predicted Higgs boson decay modes have very low branching ratios (BR) \cite{Heinemeyer:2013tqa}, and are not 
observed until now. It is not expected that with the current datasets collected with the 
ATLAS detector \cite{Aad:2008zzm} there is sensitivity to all of the predicted Higgs boson decay modes. 
However, it is still 
interesting to 
probe them in order to verify if there is an excess in any of them as it would be clear indication of new 
physics. 
Existing measurements constrain the
non-SM or "exotic" branching ratio of the Higgs boson decays to less than approximately 30\% at 
95\% confidence level (CL). 
Some Beyond Standard Model (BSM) theories predict new decay modes of the Higgs 
boson in addition to those predicted by the SM. 
%Those include models with an extended Higgs sector 
%(like 2HDM and NMSSM), Higgs Portal models of dark matter and theories of Neutral Naturalness. 
Exotic Higgs boson decays include new light resonances, as well as flavour violating decays 
and invisible decays.

\section{Search for Higgs boson decays to $\phi\gamma$}
\label{sec:phigamma}

\begin{figure}[htb]
\centering
\includegraphics[height=2in]{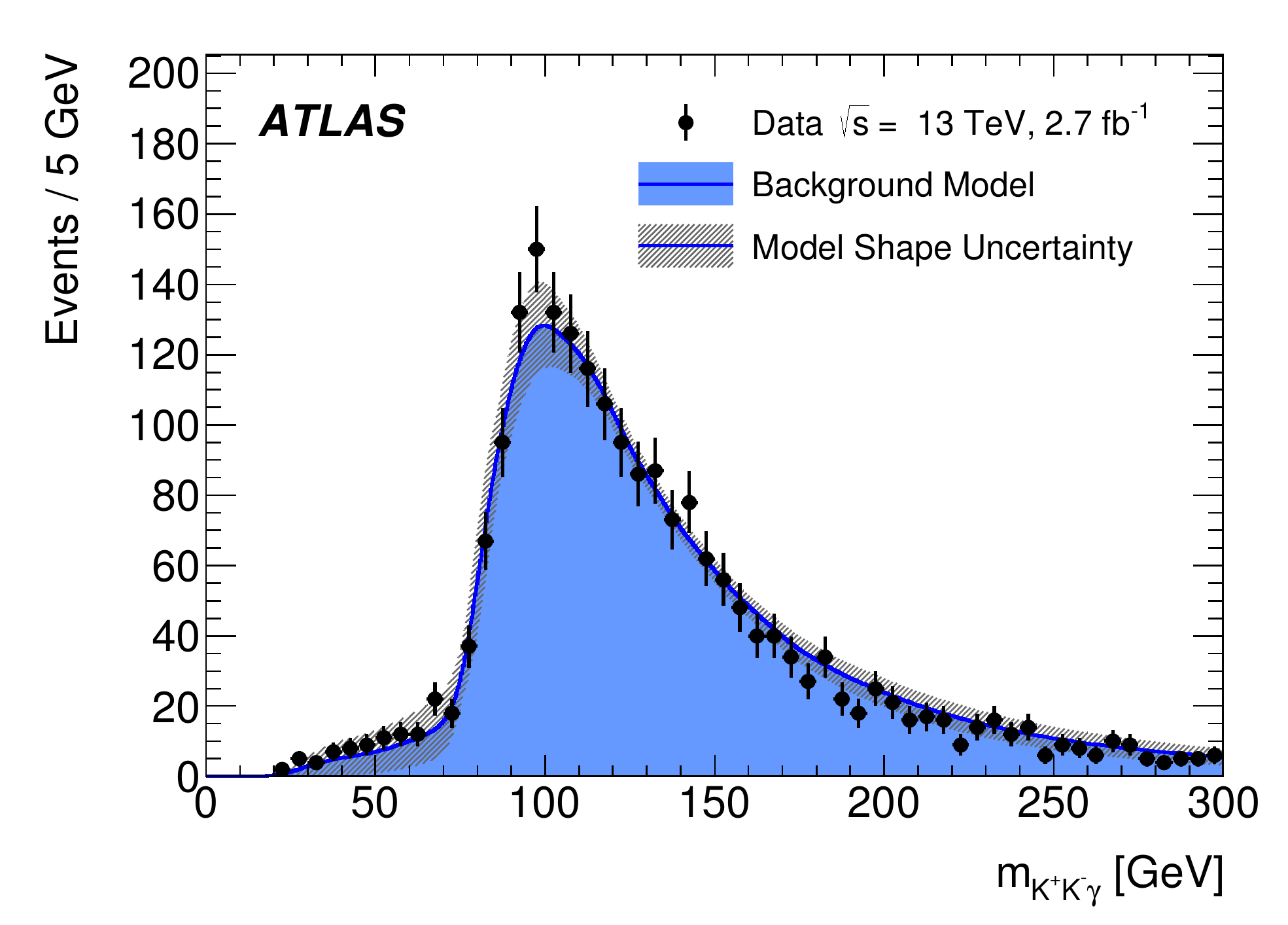}
\includegraphics[height=1.9in]{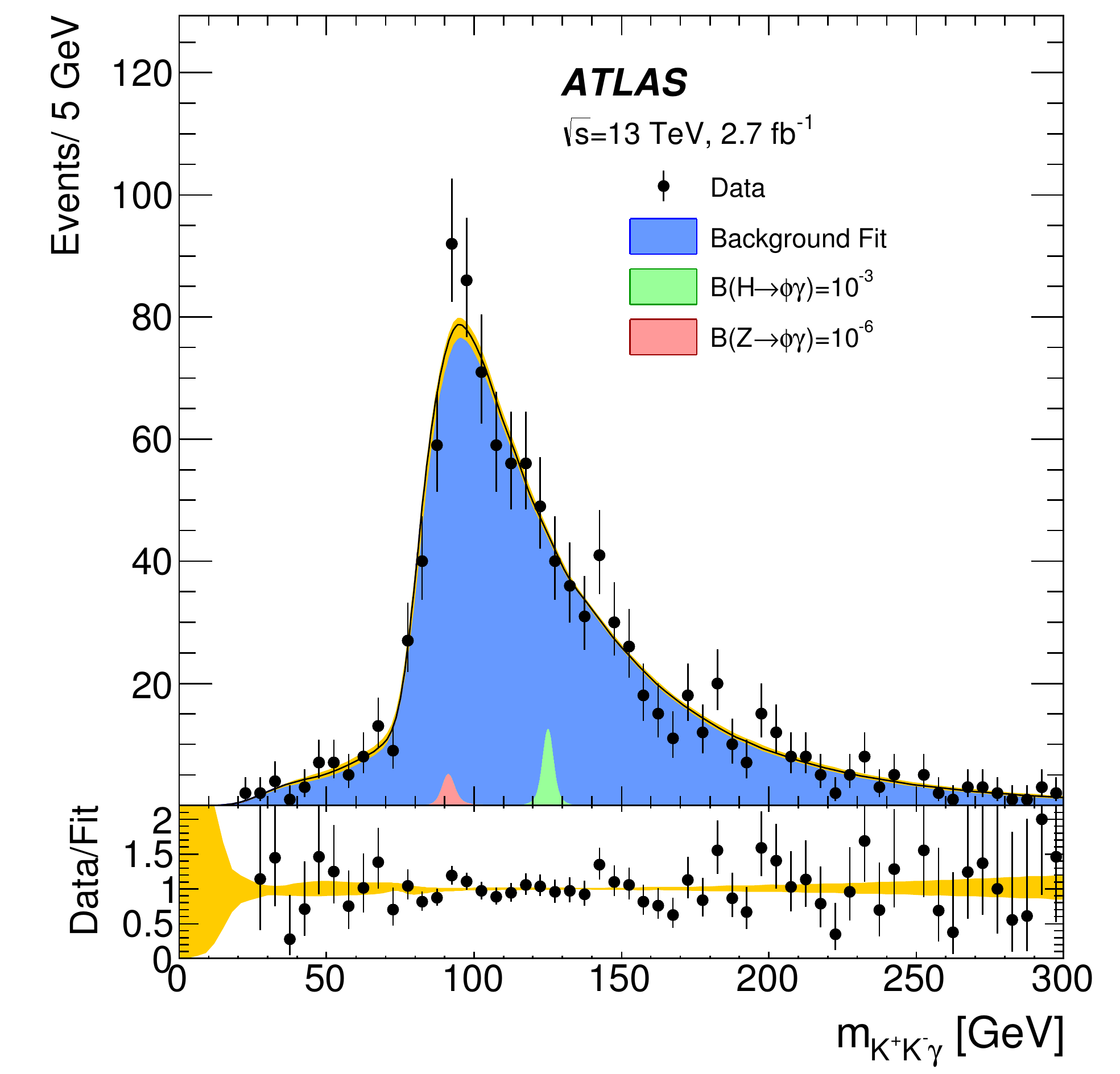}
\caption{ The distribution of $m_{K^+K^-\gamma}$ in data compared to the prediction of the 
background model for a validation control sample (on the left). The $m_{K^+K^-\gamma}$ 
distributions of the selected $\phi\gamma$ candidates, along with the results of the maximum-
likelihood fit with background-only model (on the right)~\cite{Aaboud:2016rug}.
% The Higgs and Z boson contributions, expected for branching fraction values of $10^{-3}$ and $10^{-6}$, respectively, are also shown.
}
\label{fig:figure1}
\end{figure}

A search \cite{Aaboud:2016rug} for the decays of the Higgs boson to a $\phi$ meson and a photon is 
performed with a 
$pp$ collision data sample corresponding to an integrated luminosity of 2.7 fb$^{-1}$ collected at 
$\sqrt{s}$=13 TeV. The expected SM branching fraction is
$B(H \to \phi\gamma) = (2.3\pm 0.1) \times 10^{-6}$ \cite{Koenig:2015pha}, and there is no direct 
experimental evidence about this decay mode currently. The decay
$\phi \to K^+ K^-$ is reconstructed
from pairs of oppositely charged inner detector
tracks. 
%Photons are reconstructed from clusters of energy 
%in the electromagnetic calorimeter. 
%Main backgrounds are the multijet background and $\gamma$+jets. 
The background shape is generated from the templated final state particles kinematics and 
correlations. 
%A sample of
%around 4000 $K^+K^-\gamma$ candidate events passing all of the
%kinematic selection requirements, 
%except that the photon and $\phi\to K^+K^-$ candidates are not
%required to satisfy the nominal isolation requirements (generation selection) is used determine the 
%template. 
This
background model is validated with data in samples with relaxed kinematic and 
isolation criteria. Figure \ref{fig:figure1} on the left shows the distribution of $m_{K^+K^-\gamma}$ in data 
compared to the prediction of the 
background model for a validation control sample
% defined by the generation selection with the addition of the   
%nominal photon isolation requirement. 
%The background model is normalized to the observed number 
%of events within the region shown.
% The uncertainty band corresponds to the uncertainty envelope 
%derived from variations in the background modeling procedure. 

The $m_{K^+K^-\gamma}$ distributions of the selected $\phi\gamma$ candidates, along with the 
results of the maximum-likelihood fit with background-only model is shown in Figure \ref{fig:figure1} 
on the right. 
No significant excess of events is observed above the background, and 95\% confidence level upper 
limit on the branching fraction of the Higgs boson decay to $\phi\gamma$ of $1.4\times 10^{-3}$ 
is obtained (600 times the expected SM branching fraction).

\section{A search for the decays of the Higgs boson to $J/\Psi\gamma$ and $\Upsilon(nS)\gamma$}

\begin{figure}[htb]
\centering
\includegraphics[height=1.7in]{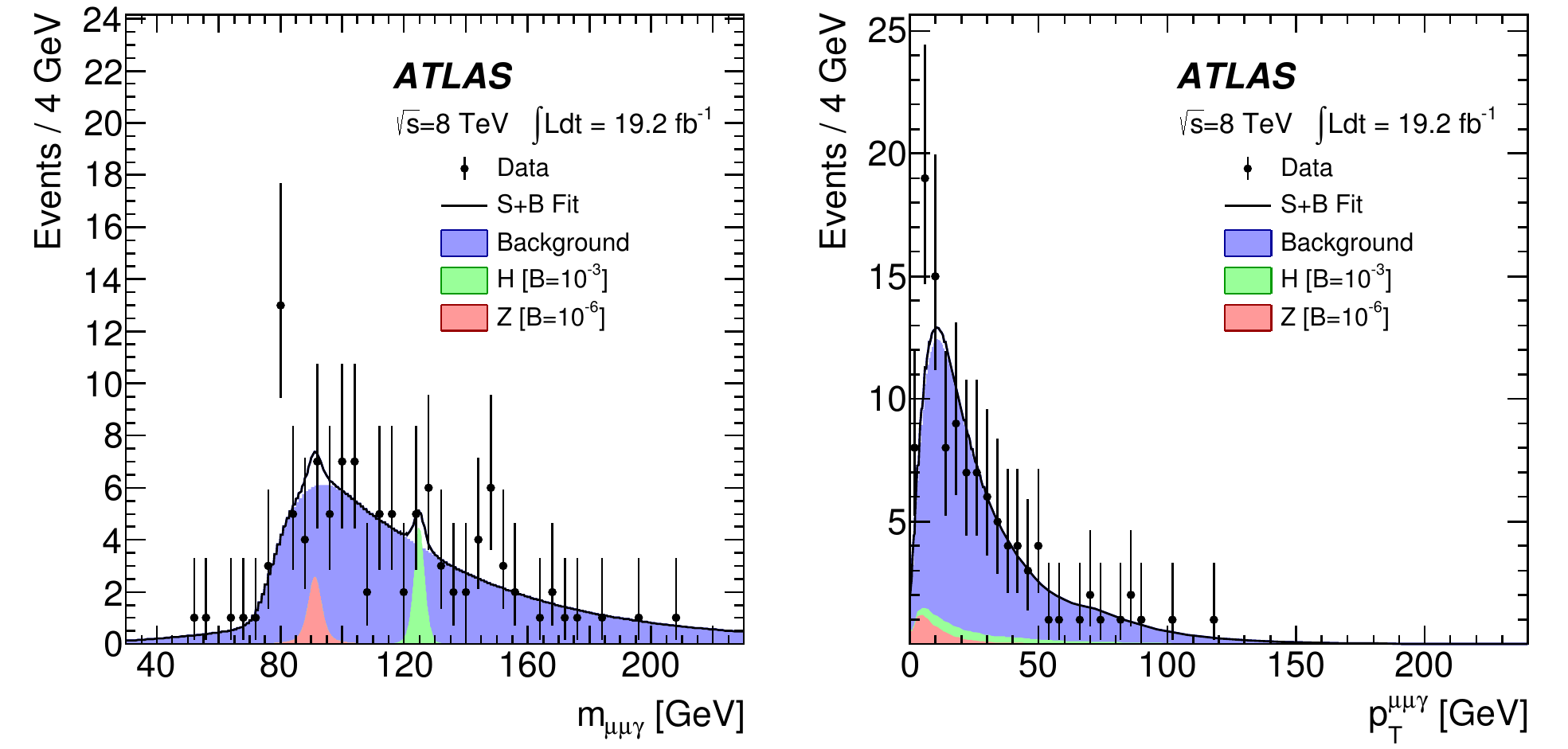}
\includegraphics[height=1.7in]{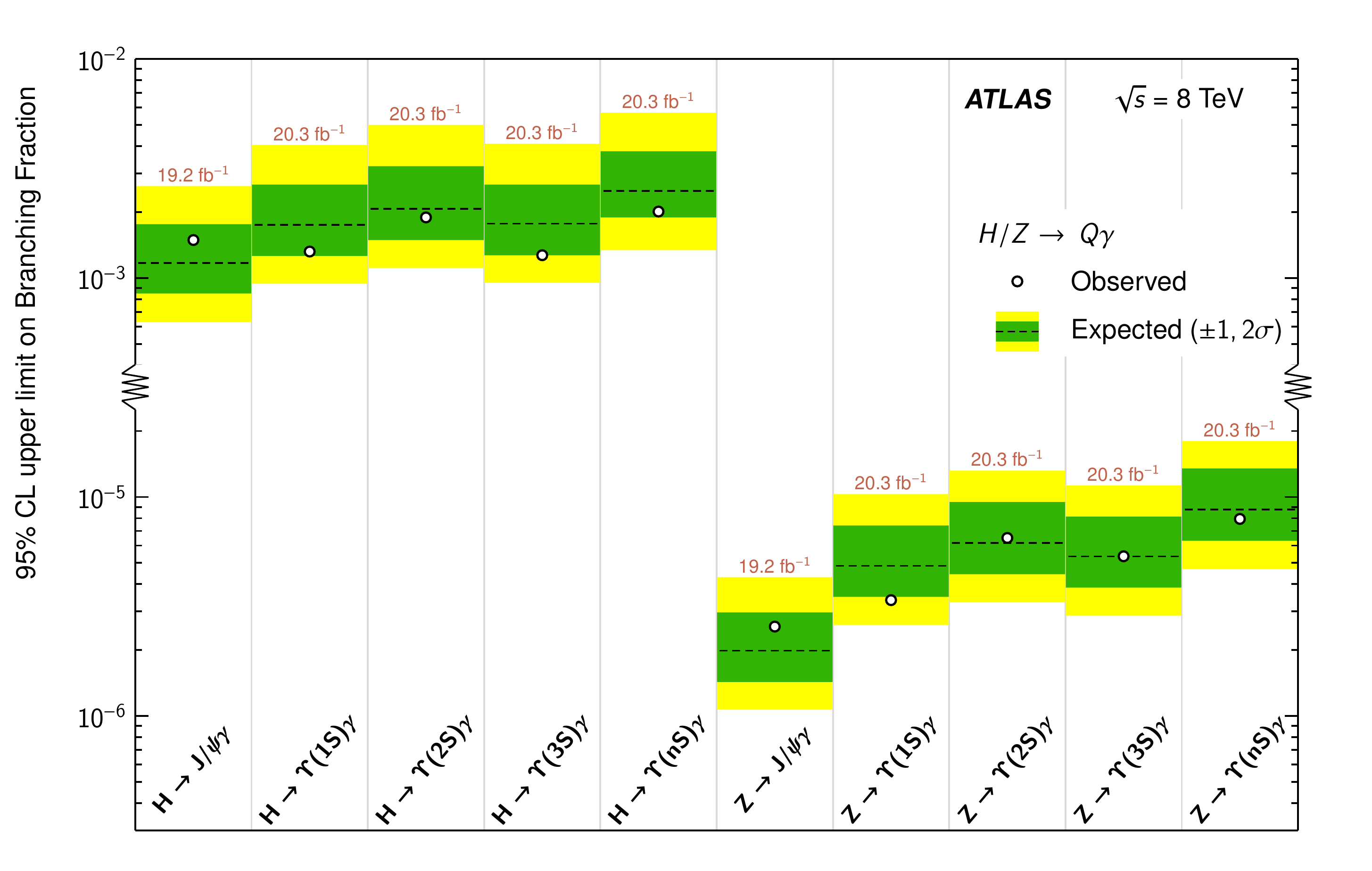}
\caption{The $m_{\mu\mu\gamma}$ and $p_T^{\mu\mu\gamma}$ distributions of the selected $J/\psi \gamma$ candidates, along with the results of the unbinned maximum likelihood fit to the signal and background model are shown on the left and in the middle. 
%The Higgs and Z boson contributions as expected for branching fraction values of $10^{-3}$ and $10^{-6}$, respectively, are also presented. 
Summary of the expected and observed branching fraction limits in the various channels studied is shown on the right~\cite{Aad:2015sda}.}
\label{fig:figure2}
\end{figure}

Rare decays of the Higgs boson to a quarkonium state and a photon may offer
unique sensitivity to both the magnitude and sign of the
Yukawa couplings of the Higgs boson to quarks. 
The expected SM branching fractions for these decays have
been calculated to be $B (H \to J/\Psi\gamma) = (2.8 \pm 0.2) \times 10^{-6}$
and $B (H\to \Upsilon(nS) \gamma$  (n=1,2,3) ) = $6.1^{+17.4}_{-6.1}, 2.0^{+1.9}_{-1.3}, 2.4^{+1.8}_{-1.3}10^{-10}$ \cite{Bodwin:2014bpa}.
No experimental evidence of these decays exists.
A search~\cite{Aad:2015sda} for the decays of the Higgs boson to $J/\Psi\gamma$ and $\Upsilon(nS)\gamma$ 
is performed with $pp$ collision data samples corresponding to integrated luminosity of up to 20.3 
fb$^{-1}$ collected at $\sqrt{s}$=8 TeV. The decays $J/\Psi \to \mu^+ \mu^-$ and
$\Upsilon(nS) \to \mu^+\mu^-$ are used to reconstruct the quarkonium
states.
% Muons are reconstructed from inner-detector tracks
%combined with independent muon spectrometer tracks or
%track segments. Photon reconstruction is seeded by clusters of energy in the electromagnetic 
%calorimeter. 
%The main source of background is the multijet background.
% dominated by inclusive 
%quarkonium production where a jet in the event is reconstructed as a photon.
The main background from inclusive multijet processes is modeled
with a non-parametric data-driven approach using
templates to describe the kinematic distributions as described in Section \ref{sec:phigamma}.
Figure \ref{fig:figure2} on the left presents the $m_{\mu\mu\gamma}$ and $p_T^{\mu\mu\gamma}$ distributions of the 
selected $J/\psi \gamma$ candidates, along with the results of the unbinned maximum likelihood fit 
to the signal and background model.
 
No significant excess of events is observed above expected 
backgrounds and 95\% C.L. upper limits are placed on the branching fractions as shown in Figure 
\ref{fig:figure2} on the right. 
In the $J/\Psi\gamma$ final state the limits are $1.5\times10^{-3}$ for the Higgs boson decays, 
while in the $\Upsilon(1S,2S,3S)\gamma$ final states the limits are $(1.3,1.9,1.3)\times10^{-3}$ 
respectively.

%REPLACE THE TEXT, FIGURE and TABLE.

%Observation of the Higgs Boson,  \cite{Aad:2012tfa},\cite{Chatrchyan:2012ufa}. 

%%%%%%%%%%%%%%%%%%%%%%%%%%%%%%%%%%%%%%%%%%%%%%%%%%%%%%%%%%%%%%%%%%%%%%%%%
%%
%%   use this format to include an .eps figure into your paper
%%

%%%%%%%%%%%%%%%%%%%%%%%%%%%%%%%%%%%%%%%%%%%%%%%%%%%%%%%%%%%%%%%%%%%%%%%%%%%

%See Figure \ref{fig:figure1} and Table \ref{tab:table1}. 

%%%%%%%%%%%%%%%%%%%%%%%%%%%%%%%%%%%%%%%%%%%%%%%%%%%%%%%%%%%%%%%%%%%%%%%%%
%%
%%   use this format to include a LaTeX table  into your paper
%%
%\begin{table}[t]
%\begin{center}
%\begin{tabular}{l|ccc}  
%Patient &  Initial level($\mu$g/cc) &  w. Magnet &  
%w. Magnet and Sound \\ \hline
 %Guglielmo B.  &   0.12     &     0.10      &     0.001  \\
 %Ferrando di N. &  0.15     &     0.11      &  $< 0.0005$ \\ \hline
%\end{tabular}
%\caption{ place the caption here }
%\label{tab:table1}
%\end{center}
%\end{table}
%%%%%%%%%%%%%%%%%%%%%%%%%%%%%%%%%%%%%%%%%%%%%%%%%%%%%%%%%%%%%%%%%%%%%%%%%%%

\section{Search for the $Z\gamma$ decay mode of the Higgs boson}

\begin{figure}[htb]
\centering
\includegraphics[height=2in]{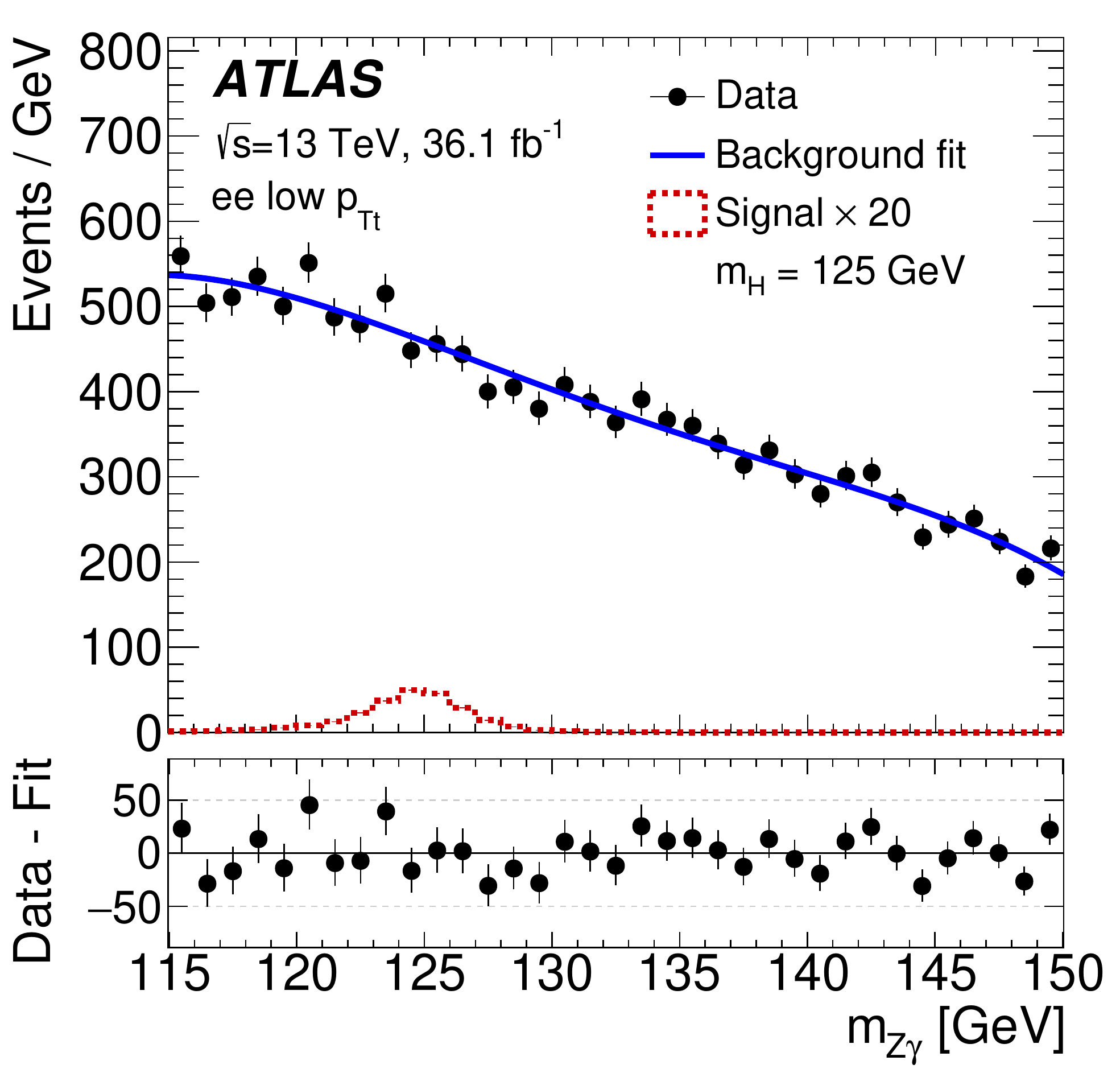}
\includegraphics[height=2in]{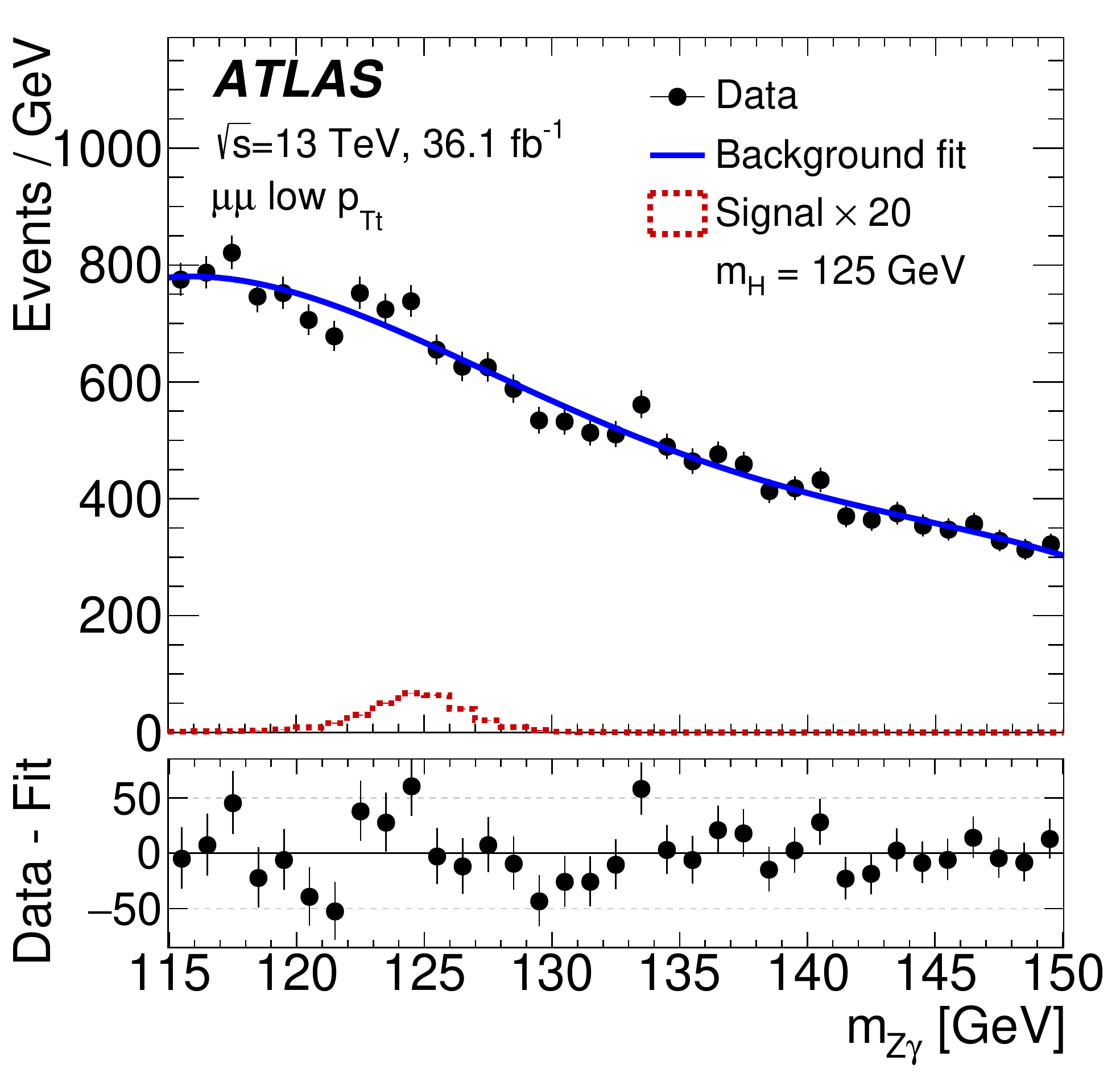}
\caption{The invariant $Z\gamma$ mass ($m_{Z\gamma}$) distributions of events satisfying the 
$H \to Z\gamma$ selection in data for the $ee$ and $\mu\mu$ channels for the low
$p_{Tt}$ category~\cite{Aaboud:2017uhw}. 
%The points represent the data and the statistical uncertainty. The solid lines show the background-only fits to
%the data, performed independently in each category. 
%The dashed histogram corresponds to the expected signal for a
%SM Higgs boson
%% with $m_H$ = 125 GeV 
%decaying to $Z\gamma$ with a rate 20 times the SM prediction. 
%The bottom part of
%the figures shows the residuals of the data with respect to the background-only fit.
}
\label{fig:figure3}
\end{figure}

This analysis~\cite{Aaboud:2017uhw} searches for the $Z\gamma$ decay of the Higgs boson exploiting $Z$ boson 
decays to pairs of electrons or muons. It uses 36.1 fb$^{-1}$ of $pp$ collisions at 
$\sqrt{s}$=13 TeV. 
The branching ratio for the Higgs boson decay to $Z\gamma$ is predicted by the SM to be 
$B(H \to Z\gamma) = (1.54 \pm 0.09) \times 10^{-3}$ for a Higgs boson mass of 125.09~GeV. 
%The $Z(\to ll)\gamma$ final
%state can be reconstructed completely and with high efficiency, good invariant mass resolution, and 
%relatively small backgrounds. 
%The main background is the non-resonant production of $Z$ bosons in 
%conjunction with photons, which is modelled with 
%simulated events.
Events are split into 6 exclusive event categories which are optimised to improve the sensitivity of 
the search and show 20\% improvement in 
sensitivity with respect to the Run1 categories. 
Figure \ref{fig:figure3} shows the invariant mass distributions $m_{Z\gamma}$ for the 
$ee$ and $\mu\mu$ channels which 
are displayed with the background-only fit performed in the range of 
$115 < m_{Z\gamma} < 150$ GeV. The variable $p_{Tt}$ is the orthogonal component of the transverse momentum 
of the $Z$ system when projected onto the axis given by the difference of the 3-momenta of the $Z$ 
boson and the photon candidate. 
No evidence of a localised excess is visible near the anticipated Higgs mass. 

The observed p-value is 0.16 under the background-only hypothesis, in which the 
dominant contribution comes from the $\mu\mu$ low-p$_{Tt}$ category.
The observed (expected - 
assuming SM $pp\to H\to Z\gamma$ production and decay) upper limit on the production cross 
section times the branching ratio for $pp\to H\to Z\gamma$ is 6.6 (5.2) times the SM prediction at the 
95\% confidence level for a Higgs boson mass of 125.09~GeV.

\section{Search for the Higgs boson produced in association with a $W$ boson and decaying to four $b$-quarks via two spin-zero particles}

A dedicated search~\cite{Aaboud:2016oyb} for exotic decays of the Higgs boson to a pair of new spin-zero particles,  
$H\to aa$, where the particle $a$ decays to $b$-quarks 
%and has a mass in the range of 20?60 GeV
is performed with the full dataset of $pp$ collisions at  $\sqrt{s}=$13~TeV  recorded in 2015, 
corresponding to an integrated luminosity of 3.2 fb$^{-1}$. The decay channel $a\to b\bar{b}$ is the preferred one 
when $m(a) > 2m(b)$.
The search is performed in events where the Higgs boson is produced in association with a $W$  
boson, giving rise to a signature of a lepton (electron or muon), missing transverse momentum ($E_T^{\mathrm{miss}}$), and multiple jets from $b$-quark decays.
% The $a$-boson can be either a scalar or a pseudoscalar under 
%parity transformations, since the decay mode considered in this search is not sensitive to the 
%difference in coupling. 
The $WH$ process is chosen because the charged lepton in the final state allows to efficiently trigger 
and identify these events against the background process of strong production of four 
$b$-jets. The analysis uses several kinematic variables combined in a multivariate
discriminant in signal regions.
% and uses control regions to reduce the uncertainties in the 
%backgrounds.

\begin{figure}[htb]
\centering
\includegraphics[height=2in]{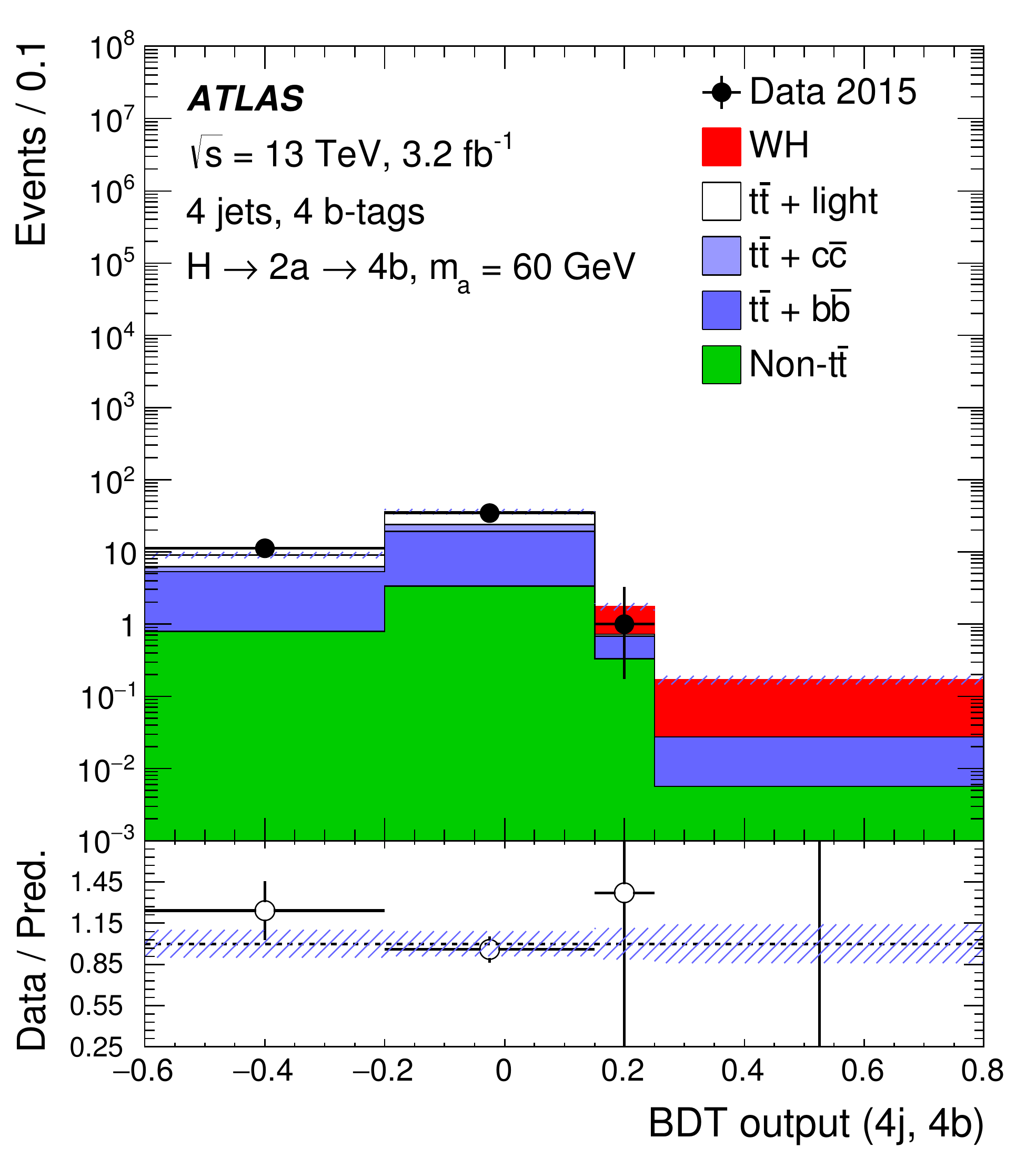}
\includegraphics[height=2in]{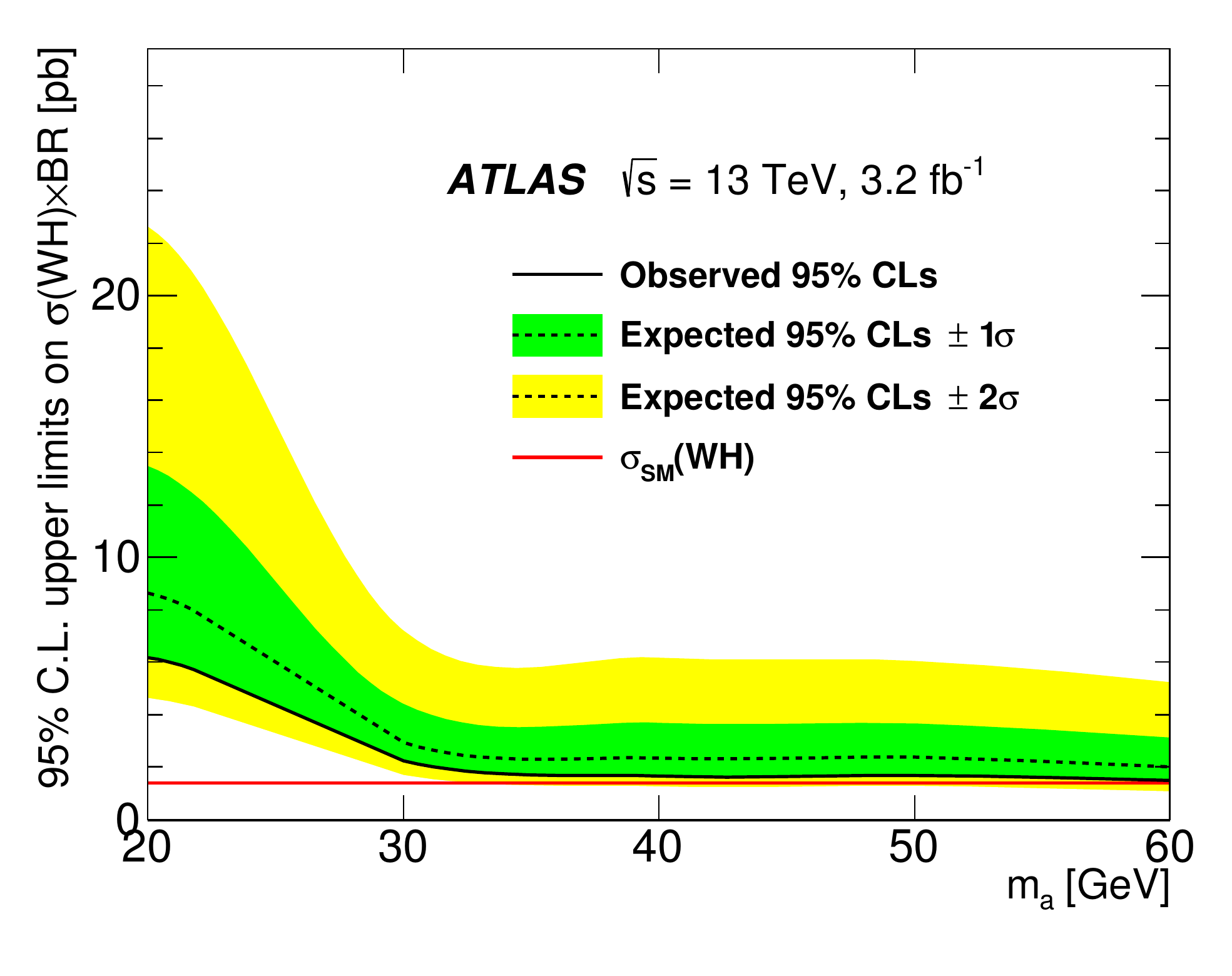}
\caption{Comparison between the data and prediction for the distribution of the BDT discriminant used in the signal regions after the fit is performed on data under the background-only hypothesis (left). 
%The hashed area represents the total uncertainty in the background. The distributions for the signal model ($WH$, $H\to 2a \to 4b$), with $m_a$ = 60 GeV, are normalised to the SM $pp \to WH$ cross section, assuming $BR(H\to aa) \times BR(a \to bb)^2$ = 1. 
Upper limit at 95\% CL on $\sigma(WH)\times BR$, where $BR=B(H\to aa) \times B(a \to bb)^2$, 
as a function of $m_a$~\cite{Aaboud:2016oyb}. 
%The observed (CLs) values (solid black line) are compared to the expected (median) (CLs) values under the background-only hypothesis (dotted black line) is shown on the right.
% The surrounding shaded bands correspond to the 68\% and 95\% CL intervals around the expected (CLs) values, denoted by $\pm 1\sigma$ and $\pm 2\sigma$, respectively. The solid red line indicates the SM $pp\to WH$ cross section, assuming $BR(H\to aa) \times BR(a \to bb)^2 = 1$.
}
\label{fig:figure4}
\end{figure}

The best fit of the background predictions to data in the binned maximum-likelihood fit is shown in
Figure \ref{fig:figure4}.
No significant excess of events above the SM prediction is 
observed, and a  95\% confidence-level upper limit is derived for the product of the production cross 
section for  $pp\to WH$  times the branching ratio for the decay  $H\to aa\to 4b$.
% The upper limit 
%ranges from 6.2 pb for an $a$-boson mass  $m_a$=20GeV  to 1.5 pb for $m_a$=60GeV.  
Assuming 
the SM $pp \to WH$ cross section, it is not possible to set limits on the branching fraction with the 
amount of data used.
% The reduced sensitivity for the light $a$-boson hypothesis is due to a lower 
%acceptance caused by overlapping $b$-jets.

\section{Search for lepton-flavour-violating decays of the Higgs boson}

Direct searches~\cite{Aad:2016blu} for lepton flavour violation (LFV) in decay of the Higgs boson to $H\to e\tau$ and 
$H\to \mu\tau$ are performed based on the data sample of $pp$ collisions corresponding to an 
integrated luminosity of 20.3 fb$^{-1}$ at a centre-of-mass energy of $\sqrt{s}=8$ TeV. The first study is
a search for $H \to e\tau$ decays in the final state with one electron and one hadronically decaying $\tau$-lepton, $\tau_{\mathrm{had}}$. The second analysis is a simultaneous search for the LFV $H \to e\tau$ and $H \to \mu\tau$ decays in the final
state with a leptonically decaying $\tau$-lepton, $\tau_{\mathrm{lep}}$. A combination of results of the earlier ATLAS search for
the LFV $H \to \mu\tau_{\mathrm{had}}$ decays and the two searches described here is also presented.
The LFV signal is searched by fitting $m_{\mathrm{MMC}}$ (for $H\to e\tau_{had}$) and $m_{\mathrm{coll}}$ (for $H\to e\tau_{\mathrm{lep}}$ and $H\to \mu\tau_{\mathrm{had}}$). 
%:Aim to reconstruct the Higgs mass
Missing mass calculator (MMC) is a version of the collinear approximation where 
relative orientations of the neutrino and other $\tau$-lepton decay products are chosen to be consistent with the mass and kinematics of a $\tau$-lepton decay. 
%Main Backgrounds are W+jets, $Z \to \tau\tau$ and $Z \to ee$ backgrounds in the
%signal regions to constrain some of the systematic uncertainties.

\begin{figure}[htb]
\centering
\includegraphics[height=2in]{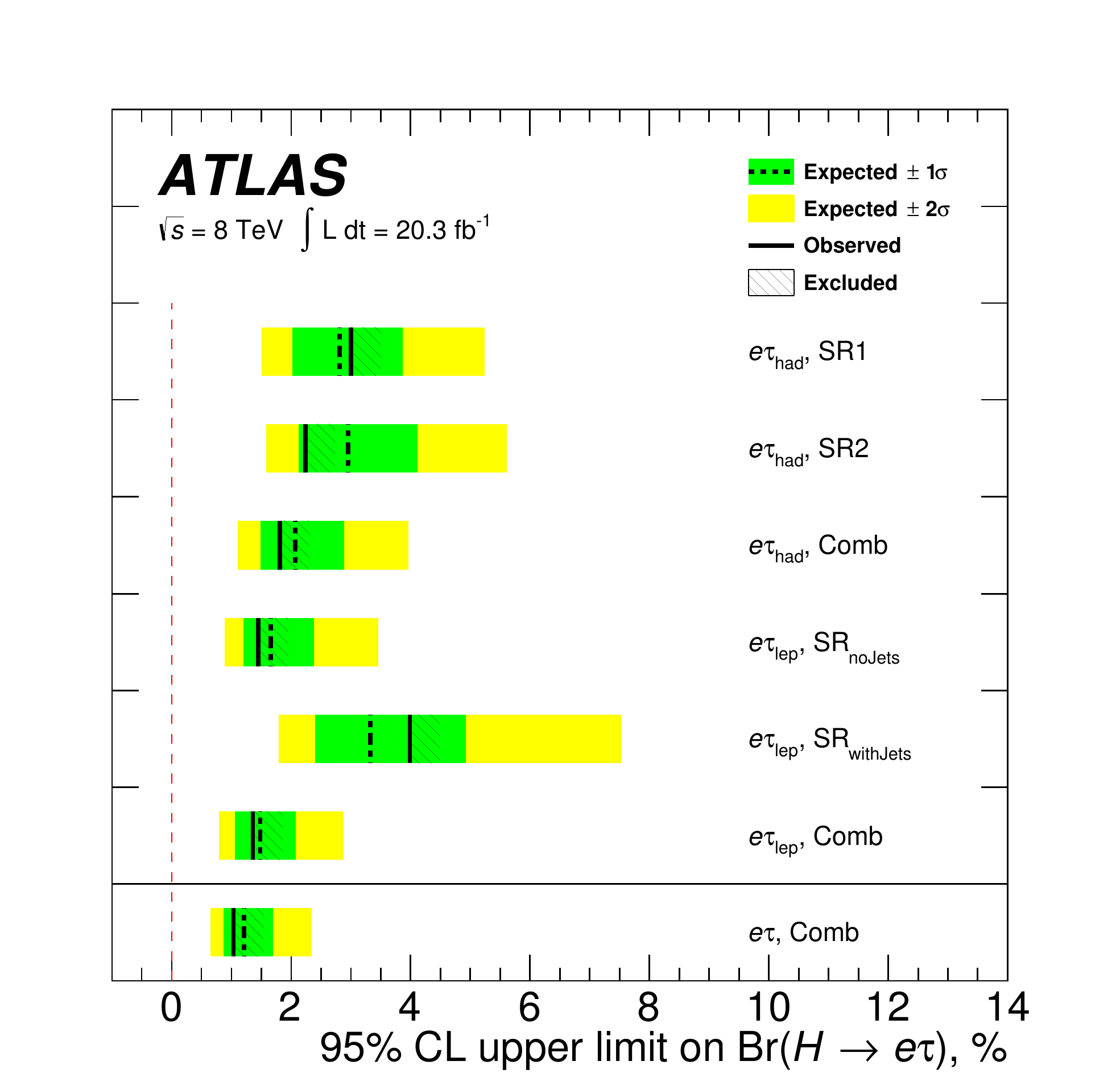}
\includegraphics[height=2in]{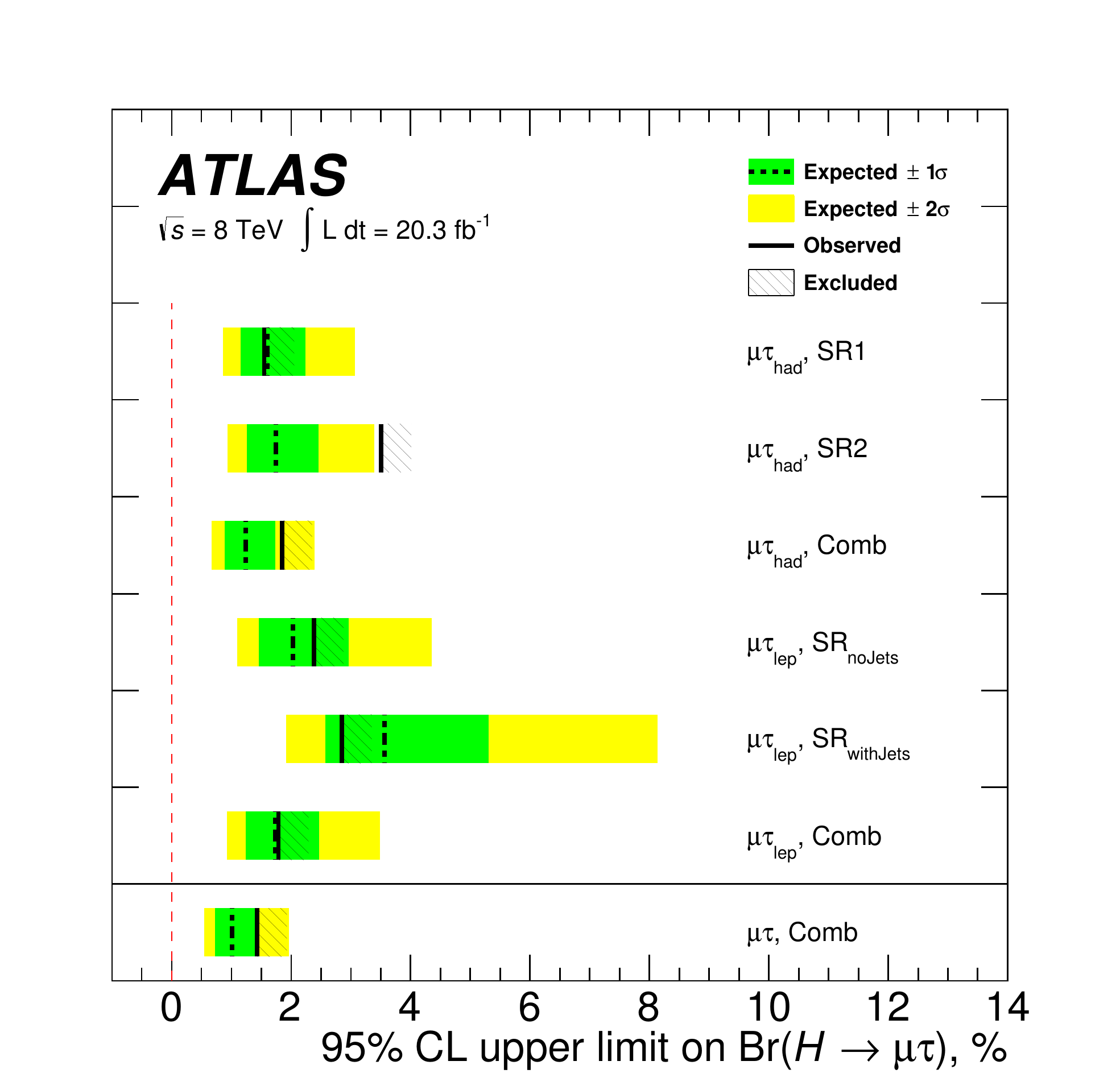}
\caption{Upper limits on LFV decays of the Higgs boson in the $H \to e\tau$ hypothesis (left) and 
$H \to \mu\tau$ hypothesis (right). The limits are computed under the assumption that either $B(H\to \mu\tau)=0$ or $B(H\to e\tau)=0$~\cite{Aad:2016blu}. The $\mu \tau_{had}$ channel is from \cite{Aad:2015gha}.}
\label{fig:figure5}
\end{figure}

%A simultaneous binned maximum-likelihood fit is performed on the $m_{MMC}$
%$e\tau$ distributions in signal and control regions to extract the LFV branching ratio BR($H \to e\tau$). 
%The fit exploits
%the control regions and the distinct shapes of the backgrounds in the
%signal regions to constrain some of the systematic uncertainties in order to increase the sensitivity 
%of the analysis.
No significant excess is observed, and upper limits on the LFV branching ratios 
are set at the 95 \% confidence level: $B (H\to e\tau)<1.04\%$ , $B (H\to\mu\tau)<1.43\%$ as shown in Figure \ref{fig:figure5}. 
% An upper limit on the LFV branching ratio BR($H\to)e\tau$)
%for a Higgs boson with $m_H$ = 125 GeV is set.
% using the CLs modified frequentist formalism with the
%test statistic based on the profile likelihood ratio. 
%The observed and the median expected 95\% CL
%upper limits are 1.81\% and 2.07+0.82
%-0.58\%, respectively. 
%Table 6 provides a summary of all results, including
%the results of the ATLAS search for the LFV $H \to \mu\tau$ decays \cite{Aad:2015gha}.

\section{Search for new phenomena in the $Z(\to ll) + E_T^{\mathrm{miss}}$ final state }

A study~\cite{ATLAS-CONF-2016-056} of the $ll+E_T^{\mathrm{miss}}$ ($l=e,\mu$) final state is performed using 13.3 fb$^{-1}$ of 13 TeV $pp$ 
collision data in 2015 and the first half of 2016. The analysis 
%is optimised to address three searches: 
%1) the search for new heavy resonances decaying to $ZZ\to ll\nu\nu$, 2) the search for dark matter in 
%association with a leptonically decaying $ZZ$ boson and 3) 
searches for an invisibly decaying Higgs boson in the channel $ZH$, $Z\to ll$, H($\to$ invisible). 
The SM Higgs boson with a measured mass of $m_H$ = 125.09 $\pm$
0.21(stat) $\pm$ 0.11(syst) GeV is predicted to have a small BF to invisible particles, $\sim$ 0.1\% in the
$H \to ZZ \to \nu\nu\nu\nu$ channel \cite{Heinemeyer:2013tqa}, which is far below the experimental sensitivity of the current analyses.

\begin{figure}[htb]
\centering
\includegraphics[height=2in]{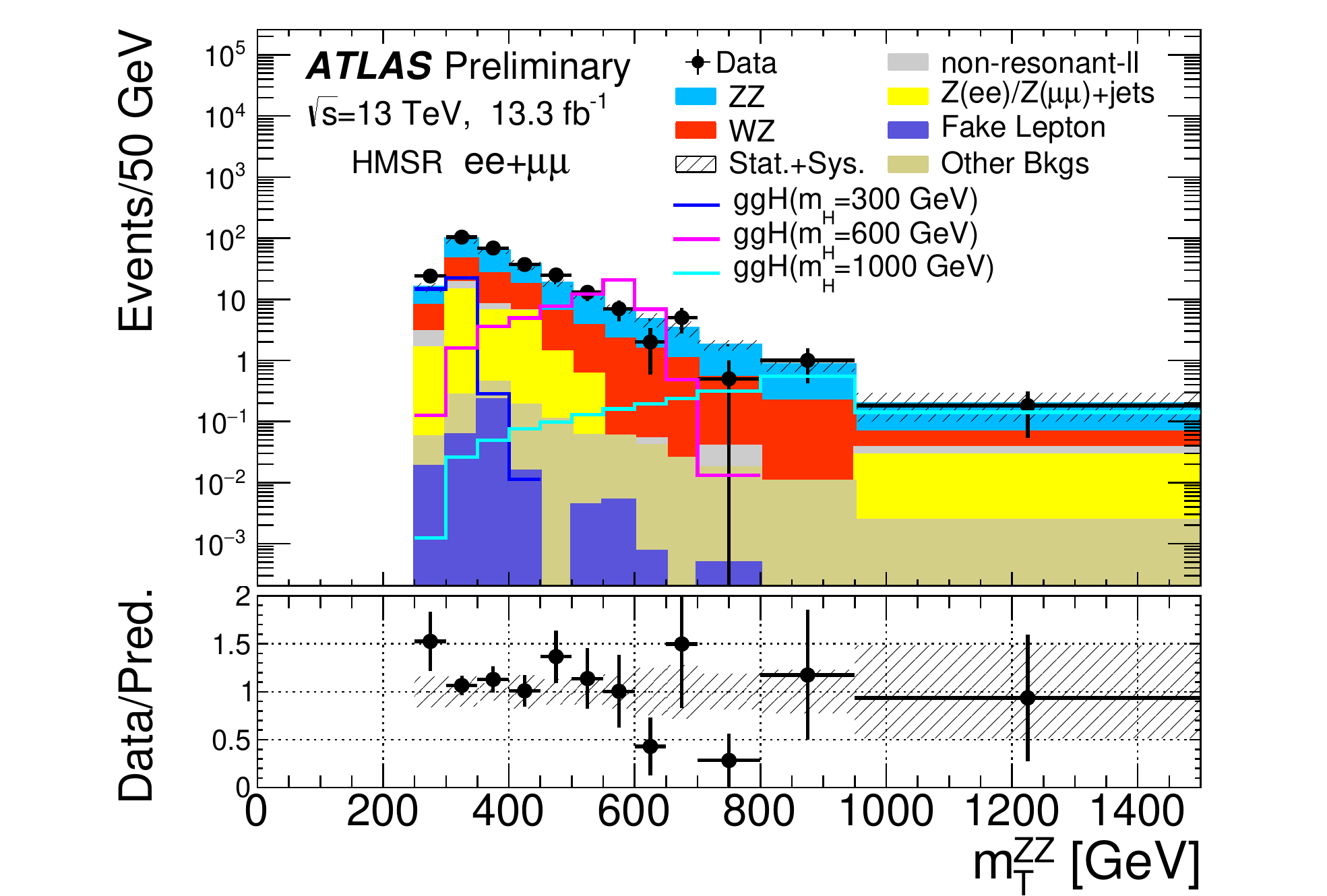}
\includegraphics[height=2in]{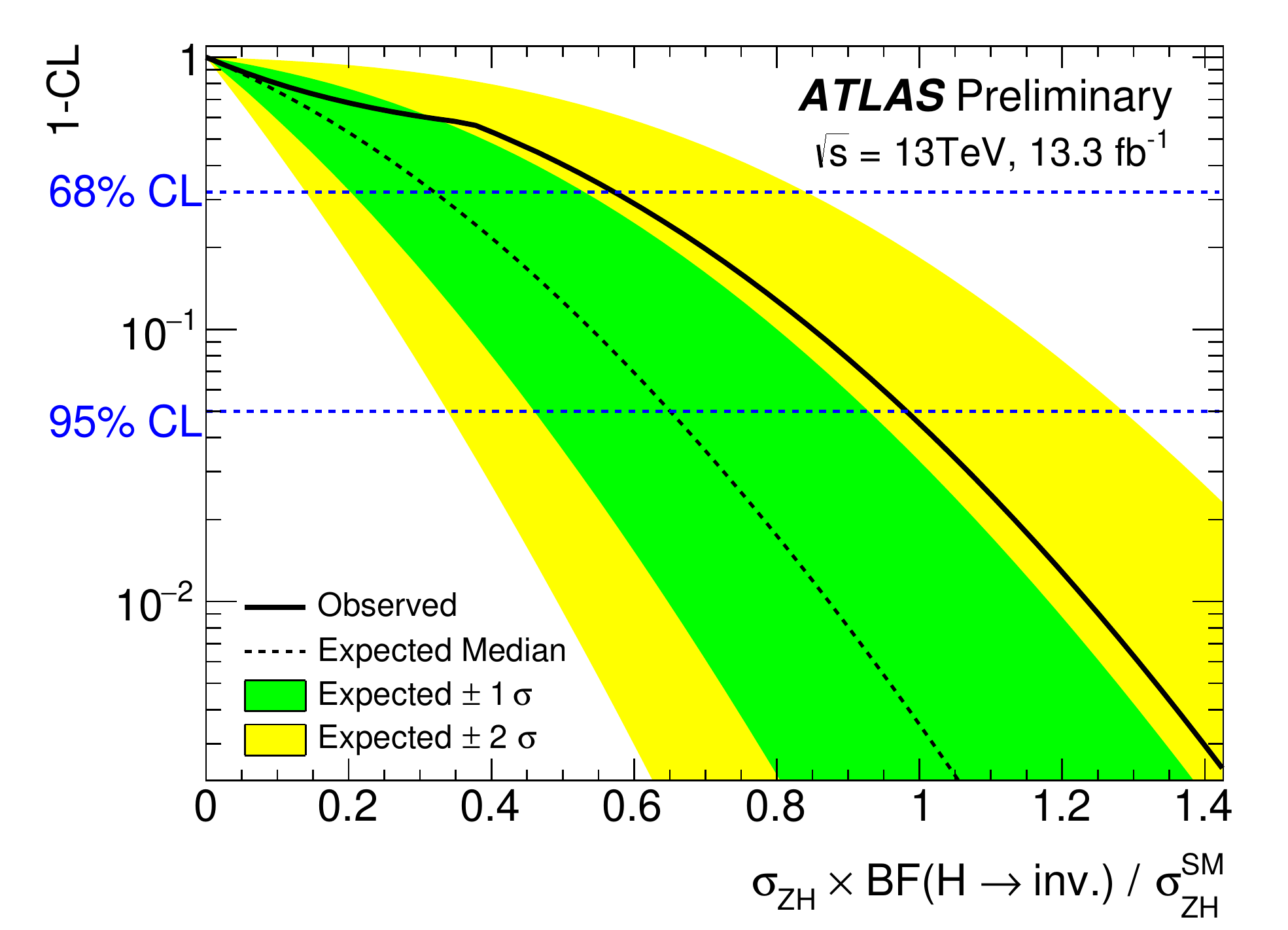}
\caption{$m_T^{ZZ}$ distributions in one signal region for the combined $ee+\mu\mu$  
channels (left). The stacked histograms represent the background predictions, while the blue, pink and 
cyan curves give the predicted signal distributions for a heavy Higgs boson with different masses. 
%The total uncertainty of the background expectation is shown in the grey shaded band.
% The 
%number of entries in each bin corresponds to the number of events per 50 GeV in that region.
Confidence levels corresponding to upper limits as a function of $\sigma_{\mathrm{ZH}} \times B(H\to \mathrm{inv.} )$ / $\sigma_{\mathrm{ZH}}^{\mathrm{SM}}$~\cite{ATLAS-CONF-2016-056}.
% scanned from 0 to 1.4 (right). 
%The expected and observed confidence levels are shown as the dashed black and solid black lines, respectively. 
%The green and yellow bands give the $\pm 1\sigma$ and $\pm 2\sigma$ uncertainties of the 
%expected confidence levels, respectively. The shown confidence levels can be interpreted as those 
%on the upper limits of $BF(H(\to invisible))$, for the region with the x-axis value less than one. The 
%95\% CL upper limit on $BF(H(\to invisible))$ can be read from the crossing points between the 
%dashed blue " 95\% CL'' line and the respective confidence level curve. 
}
\label{fig:figure6}
\end{figure}

New physics is searched for as an excess over the 
SM predictions in the $ZZ$ transverse mass $m^{ZZ}_T$ (Eq.~\ref{eq:mZZ}) 
\begin{equation}
\label{eq:mZZ}
(m^{ZZ}_T)^2 = \bigg(\sqrt {m^2_Z + \big| p_T^{ll} \big|^2} + \sqrt {m_Z^2+\big|E_T^{\mathrm{miss}}\big|^2}\bigg)^2 - \Big| \vec{p}_T^{ll} + \vec{E}_T^{\mathrm{miss}}\Big|^2
\end{equation}
distribution and in the 
$E_T^{\mathrm{miss}}$ distribution. 
Figure \ref{fig:figure6} left presents the $m_T^{ZZ}$ distributions in one signal region for the combined 
$ee+\mu\mu$ channels. 
Results are found to be compatible with SM expectations. Figure \ref{fig:figure6} right presents a distribution of the confidence levels corresponding to each value of upper limits on 
$\sigma(Z(\to ll)H(\to \mathrm{invisible}))$ divided by the
SM prediction of the $ZH$ production cross-section (with $m_H$=125 GeV) scanned from 0 to 1.4. 
The shown
confidence levels can be interpreted as that on the upper limits of B($H \to$ invisible), for the region 
with
the x-axis value less than one. The expected and observed upper limit on B($H\to$ invisible) at 95\% CL
is 65\% and 98\%, respectively. 
%The observed limit is larger than the expectation, and this is caused by the
%moderate data excess in the EmissT distributions in both the ee and µµ channels

\section{Conclusions}

Searches for rare and exotic Higgs decays with the ATLAS detector with 8 TeV and 13 TeV data 
are presented. 
%An increase of sensitivity is expected with more data. 
%There are still more possibilities to explore to cover full spectrum. 
No significant deviations from the Standard Model are found. Limits on different models 
of Beyond Standard Model physics are extended or set for the first time.
There are still more possibilities to explore to cover the full spectrum. 
%%  if necessary
%\Acknowledgements
%I am grateful to XYZ for fruitful discussions.

\end{document}